\documentstyle[twoside,psfig]{article}

\catcode`\@=11
\long\def\@makefntext#1{
\protect\noindent \hbox to 3.2pt {\hskip-.9pt  
$^{{\eightrm\@thefnmark}}$\hfil}#1\hfill}		%CAN BE USED 

\def\thefootnote{\fnsymbol{footnote}}
\def\@makefnmark{\hbox to 0pt{$^{\@thefnmark}$\hss}}	%ORIGINAL 
	
\def\ps@myheadings{\let\@mkboth\@gobbletwo
\def\@oddhead{\hbox{}
\rightmark\hfil\eightrm\thepage}   
\def\@oddfoot{}\def\@evenhead{\eightrm\thepage\hfil
\leftmark\hbox{}}\def\@evenfoot{}
\def\sectionmark##1{}\def\subsectionmark##1{}}

\oddsidemargin=\evensidemargin
\renewcommand{\thefootnote}{\fnsymbol{footnote}}

\newcounter{sectionc}\newcounter{subsectionc}\newcounter{subsubsectionc}
\renewcommand{\section}[1] {\vspace{12pt}\addtocounter{sectionc}{1} 
\setcounter{subsectionc}{0}\setcounter{subsubsectionc}{0}\noindent 
	{\tenbf\thesectionc. #1}\par\vspace{5pt}}
\renewcommand{\subsection}[1] {\vspace{12pt}\addtocounter{subsectionc}{1} 
	\setcounter{subsubsectionc}{0}\noindent 
	{\bf\thesectionc.\thesubsectionc. {\kern1pt \bfit #1}}\par\vspace{5pt}}
\renewcommand{\subsubsection}[1] {\vspace{12pt}\addtocounter{subsubsectionc}{1}
	\noindent{\tenrm\thesectionc.\thesubsectionc.\thesubsubsectionc.
	{\kern1pt \tenit #1}}\par\vspace{5pt}}
\newcommand{\nonumsection}[1] {\vspace{12pt}\noindent{\tenbf #1}
	\par\vspace{5pt}}

\newcounter{appendixc}
\newcounter{subappendixc}[appendixc]
\newcounter{subsubappendixc}[subappendixc]
\renewcommand{\thesubappendixc}{\Alph{appendixc}.\arabic{subappendixc}}
\renewcommand{\thesubsubappendixc}
	{\Alph{appendixc}.\arabic{subappendixc}.\arabic{subsubappendixc}}

\renewcommand{\appendix}[1] {\vspace{12pt}
        \refstepcounter{appendixc}
        \setcounter{figure}{0}
        \setcounter{table}{0}
        \setcounter{lemma}{0}
        \setcounter{theorem}{0}
        \setcounter{corollary}{0}
        \setcounter{definition}{0}
        \setcounter{equation}{0}
        \renewcommand{\thefigure}{\Alph{appendixc}.\arabic{figure}}
        \renewcommand{\thetable}{\Alph{appendixc}.\arabic{table}}
        \renewcommand{\theappendixc}{\Alph{appendixc}}
        \renewcommand{\thelemma}{\Alph{appendixc}.\arabic{lemma}}
        \renewcommand{\thetheorem}{\Alph{appendixc}.\arabic{theorem}}
        \renewcommand{\thedefinition}{\Alph{appendixc}.\arabic{definition}}
        \renewcommand{\thecorollary}{\Alph{appendixc}.\arabic{corollary}}
        \renewcommand{\theequation}{\Alph{appendixc}.\arabic{equation}}
        \noindent{\tenbf Appendix \theappendixc #1}\par\vspace{5pt}}
\newcommand{\subappendix}[1] {\vspace{12pt}
        \refstepcounter{subappendixc}
        \noindent{\bf Appendix \thesubappendixc. {\kern1pt \bfit #1}}
	\par\vspace{5pt}}
\newcommand{\subsubappendix}[1] {\vspace{12pt}
        \refstepcounter{subsubappendixc}
        \noindent{\rm Appendix \thesubsubappendixc. {\kern1pt \tenit #1}}
	\par\vspace{5pt}}

\newcommand{\textlineskip}{\baselineskip=13pt}
\newcommand{\smalllineskip}{\baselineskip=10pt}

\def\eightcirc{
\begin{picture}(0,0)
\put(4.4,1.8){\circle{6.5}}
\end{picture}}
\def\eightcopyright{\eightcirc\kern2.7pt\hbox{\eightrm c}} 

\newcommand{\copyrightheading}[1]
	{\vspace*{-2.5cm}\smalllineskip{\flushleft
	{\footnotesize International Journal of Modern Physics A #1}\\
	{\footnotesize $\eightcopyright$\, World Scientific Publishing
	 Company}\\
	 }}

\newcommand{\publisher}[2]{{\begin{center}\footnotesize\smalllineskip 
	Received #1\\
	Revised #2
	\end{center}
	}}

\def\abstracts#1#2#3{{
	\centering{\begin{minipage}{4.5in}\footnotesize\baselineskip=10pt
	\parindent=0pt #1\par 
	\parindent=15pt #2\par
	\parindent=15pt #3
	\end{minipage}}\par}} 

\renewenvironment{thebibliography}[1]
	{\frenchspacing
	 \ninerm\baselineskip=11pt
	 \begin{list}{\arabic{enumi}.}
	{\usecounter{enumi}\setlength{\parsep}{0pt}
	 \setlength{\leftmargin 12.7pt}{\rightmargin 0pt} %FOR 1--9 ITEMS
	 \setlength{\itemsep}{0pt} \settowidth
	{\labelwidth}{#1.}\sloppy}}{\end{list}}

\newcounter{itemlistc}
\newcounter{romanlistc}
\newcounter{alphlistc}
\newcounter{arabiclistc}

\newcommand{\fcaption}[1]{
        \refstepcounter{figure}
        \setbox\@tempboxa = \hbox{\footnotesize Fig.~\thefigure. #1}
        \ifdim \wd\@tempboxa > 5in
           {\begin{center}
        \parbox{5in}{\footnotesize\smalllineskip Fig.~\thefigure. #1}
            \end{center}}
        \else
             {\begin{center}
             {\footnotesize Fig.~\thefigure. #1}
              \end{center}}
        \fi}

\newcommand{\tcaption}[1]{
        \refstepcounter{table}
        \setbox\@tempboxa = \hbox{\footnotesize Table~\thetable. #1}
        \ifdim \wd\@tempboxa > 5in
           {\begin{center}
        \parbox{5in}{\footnotesize\smalllineskip Table~\thetable. #1}
            \end{center}}
        \else
             {\begin{center}
             {\footnotesize Table~\thetable. #1}
              \end{center}}
        \fi}

\def\@citex[#1]#2{\if@filesw\immediate\write\@auxout
	{\string\citation{#2}}\fi
\def\@citea{}\@cite{\@for\@citeb:=#2\do
	{\@citea\def\@citea{,}\@ifundefined
	{b@\@citeb}{{\bf ?}\@warning
	{Citation `\@citeb' on page \thepage \space undefined}}
	{\csname b@\@citeb\endcsname}}}{#1}}

\newif\if@cghi
\def\cite{\@cghitrue\@ifnextchar [{\@tempswatrue
	\@citex}{\@tempswafalse\@citex[]}}
\def\citelow{\@cghifalse\@ifnextchar [{\@tempswatrue
	\@citex}{\@tempswafalse\@citex[]}}
\def\@cite#1#2{{$\null^{#1}$\if@tempswa\typeout
	{IJCGA warning: optional citation argument 
	ignored: `#2'} \fi}}

\def\pmb#1{\setbox0=\hbox{#1}
	\kern-.025em\copy0\kern-\wd0
	\kern.05em\copy0\kern-\wd0
	\kern-.025em\raise.0433em\box0}

\def\fnt#1#2{\footnotetext{\kern-.3em
	{$^{\mbox{\scriptsize #1}}$}{#2}}}

\def\thefootnote{\fnsymbol{footnote}}
\def\@makefnmark{\hbox to 0pt{$^{\@thefnmark}$\hss}}	%ORIGINAL 
	
\def\ps@myheadings{%
    \let\@oddfoot\@empty\let\@evenfoot\@empty
    \def\@evenhead{\slshape\leftmark\hfil}%       %EVEN PAGE
    \def\@oddhead{\hfil{\slshape\rightmark}}%     %ODD PAGE
    \let\@mkboth\@gobbletwo
    \let\sectionmark\@gobble
    \let\subsectionmark\@gobble
    }
\font\tenrm=cmr10
\font\tenit=cmti10 
\font\tenbf=cmbx10
\font\bfit=cmbxti10 at 10pt
\font\ninerm=cmr9

\font\eightrm=cmr8

\def\qed{\hbox{${\vcenter{\vbox{			%HOLLOW SQUARE
   \hrule height 0.4pt\hbox{\vrule width 0.4pt height 6pt
   \kern5pt\vrule width 0.4pt}\hrule height 0.4pt}}}$}}

\renewcommand{\thefootnote}{\fnsymbol{footnote}}  %USE SYMBOLIC FOOTNOTE

\begin{document}
\setlength{\textheight}{7.7truein}  %for 2nd page onwards

\thispagestyle{empty}

%\markboth{\protect{\footnotesize\it Instructions for Typesetting
%Manuscripts}}{\protect{\footnotesize\it Instructions for
%Typesetting Manuscripts}}

\normalsize\textlineskip

\setcounter{page}{1}

\copyrightheading{}		%{Vol.~0, No.~0 (2000) 000--000}

\vspace*{0.88truein}

%\fpage{1}
\centerline{\bf NEUTRON STARS AND THE FERMIONIC CASIMIR EFFECT}
\vspace*{0.035truein}
\vspace*{0.37truein}
\centerline{\footnotesize PIOTR MAGIERSKI}
\baselineskip=12pt
\centerline{\footnotesize\it Institute of Physics, 
Warsaw University of Technology, ul. Koszykowa 75}
\baselineskip=10pt
\centerline{\footnotesize\it Warsaw, PL--00662, Poland}
\vspace*{10pt}
\centerline{\footnotesize AUREL BULGAC}
\baselineskip=12pt
\centerline{\footnotesize\it Department of Physics,  University of Washington}
\baselineskip=10pt
\centerline{\footnotesize\it Seattle, WA 98195--1560, USA}
\vspace*{10pt}
\centerline{\footnotesize PAUL-HENRI HEENEN}
\baselineskip=12pt
\centerline{\footnotesize\it Service de Physique Nucl\'{e}aire Th\'{e}orique,
 U.L.B - C.P. 229}
\baselineskip=10pt
\centerline{\footnotesize\it Brussels, B 1050, Belgium}
\vspace*{0.225truein}
\publisher{(received date)}{(revised date)}

\vspace*{0.21truein}
\abstracts{The inner crust of neutron stars consists of nuclei
of various shapes immersed in a neutron gas and stabilized
by the Coulomb interaction in the form of a crystal lattice.
The scattering of neutrons on nuclear inhomegeneities
leads to the quantum correction to the total energy of the
system. This correction resembles the Casimir energy and
turns out to have a large influence on the structure of the crust.
}{}{}

\textlineskip			%) USE THIS MEASUREMENT WHEN THERE IS
\vspace*{12pt}			%) NO SECTION HEADING

%\vspace*{1pt}\textlineskip	%) USE THIS MEASUREMENT WHEN THERE IS
%\section{General Appearance}	%) A SECTION HEADING
\vspace*{-0.5pt}

\setcounter{footnote}{0}
\renewcommand{\thefootnote}{\alph{footnote}}

The average density of a typical neutron star is by
$3-4$ times larger than the density inside an atomic nucleus.
Even though, in the neutron star crust
the nucleon density is relatively low and do not exceed $0.1$ fm$^{-3}$.
The width of this layer reaches almost $10$\% of the star radius
and its structure is quite complex.
The outer part of the crust consists of nuclei forming
a lattice immersed in an electron gas which become relativistic
already at about $10^{6}$ g cm$^{-3}$.
Deeper in the star, in the inner crust, due to high density and pressure,
the last occupied neutron levels in nuclei are no longer bound, and
a neutron gas is formed. The structure of this region can thus be
viewed as an inhomogeneous neutron matter, where nuclei play
the role of impurities.
The electrons at these densities are ultrarelativistic particles and
therefore the electron-nucleus correlations, which are responsible for
the complications of electronic structure in atoms and solids
become unimportant. Eventually, at the bottom of the crust, the
uniform nuclear matter is formed.

The structure of the outer parts of neutron stars are important for
understanding of several observational issues.
Namely, one expects that such phenomena as: thermal X-ray emission
from the stellar surface, X-ray burst sources, or
the sudden speed-ups in the rotation rate of
some neutron stars may be related to its crustal properties.
It provides thus a motivation for theoretical studies,
and indeed the investigation of the crust structure
has been the subject of a considerable theoretical effort
(see e.g. \cite{bbp,oya2,pra,dhm,dha} and reference therein).
In particular, it has been found that in the inner crust
the interplay between the Coulomb and the surface energies
leads to the appearance of various nuclear phases,
characterized by different shapes.
An agreement has been reached on the existence of five phases which are formed
in the region where the nucleon density varies from $0.03$ to $0.1$ fm$^{-3}$.
Most theoretical calculations predict a chain of phase transitions
as the density increases: spherical nuclei $\rightarrow$
rods (``spaghetti'' phase)
$\rightarrow$ slabs (``lasagna'' phase)
$\rightarrow$ tubes $\rightarrow$ bubbles
$\rightarrow$ uniform matter.
However in all to date approaches shell effects
have been either completely neglected, as in the models based on a liquid drop
formula or Thomas-Fermi models, or have been taken into account only for bound
nucleons. Namely, the total energy density of the system can be expressed
schematically in the form:
\begin{equation}  \label{energy}
E=E_{vol}+E_{surf}+E_{Coul}+E_{shell},
\end{equation}
where the volume term $E_{vol}$
describes bulk properties of the neutron-proton-electron
(npe) matter, $E_{surf}$
denotes the surface energy of nuclei, and $E_{Coul}$
takes into account the Coulomb interaction between protons
and electrons. The last term denotes the shell energy
and was usually omitted in the previous approaches. It represents
the quantum correction to the total energy and has
two origins. The first one is associated with bound nucleons.
Since the nuclei form a lattice the single-particle
bound states form narrow bands. Their distribution
is not a smooth function of energy and therefore
gives rise to an energy correction which is termed shell energy.
The change of the total energy can be
computed quite accurately using the shell correction method,
once the single-particle spectrum is known \cite{strutin,strut}.
It was shown however
that this correction play a marginal role and does not influence the
phase transition pattern \cite{oya2}.
The second effect is related to unbound neutrons which may scatter on
inhomogeneities and form resonant states. The situation is somewhat
similar to the Casimir effect in quantum field theory and condensed
matter (see e.g. Refs. \cite{cas,fish,kgo} and references therein),
where the fluctuation induced interaction leads to an additional
energy correction to the total energy of a system.
In the same way, the scattering of unbound neutrons
leads to an effective interaction between nuclei
immersed in the neutron environment. This effect
can be termed the fermionic Casimir effect and has been studied
for simple geometries in Refs. \cite{bma1,bma2,bma3,bwi,bma4}.

The shell energy for unbound neutrons can be calculated from
the formula:
%----------------------------------------------------------------------
\begin{equation} \label{shell}
E_{shell}^{out}=
\int_{-\infty}^ \mu    d\varepsilon\varepsilon g  (\varepsilon ,{\bf l})
-\int_{-\infty}^{\mu _0}d\varepsilon\varepsilon g_0(\varepsilon ,{\bf l}).
\end{equation}
%----------------------------------------------------------------------
In the above equation $g _0(\varepsilon , {\bf l})$ stands for the
Thomas--Fermi or liquid drop density of states related to the neutron
gas and $g (\varepsilon ,{\bf l} )$ is the true quantum density of
states in the presence of inhomogeneities. An ensemble
of parameters ${\bf l}$ describes these objects and their
relative geometrical arrangement. The parameters: $\mu$ and
$\mu _0$ are determined from the requirement that the system has a
given average density of unbound neutrons:
%----------------------------------------------------------------------
\begin{equation}
\rho_{n}^{out} =
\int _{-\infty }^\mu d\varepsilon  g (\varepsilon ,{\bf l} )
=  \int _{-\infty }^{\mu _0} d\varepsilon g _0(\varepsilon ,{\bf l} ) . 
\end{equation}
%----------------------------------------------------------------------
Since in infinite matter the presence of various inhomogeneities does
not lead to the formation of discrete levels, the effects we
shall consider here arise from the scattering
states, which is in complete analogy with the procedure for computing
the Casimir energy. The only difference comes from the fact that
the integration in the formula (\ref{shell}) is performed up
to the Fermi energy instead of infinity, as in the
expression for the Casimir energy. This fact is responsible
for an oscillatory behavior of the shell energy as a function
of the distance between nuclei.

If the impurities were small compared
to the neutron Fermi wavelength then the generated interaction
would originate mainly from the s-wave scattering giving rise to
the Ruderman-Kittel correction which is small compared to the
Fermi energy of neutrons \cite{rki,fwa}.
However the s-wave scattering limit is not valid in this case
since the neutron Fermi momentum $k^{n}_{F}$ is of
the order of $1 fm$ and thus $k^{n}_{F}R > 1$ where $R$ is a size of
an inhomogeneity. Consequently the contribution of higher partial waves
to the scattering process cannot be neglected, giving rise
to much larger energy correction, as for large objects more of the incident
wave will be reflected \cite{bma1,bma2,bma3,bwi,bma4}.

\begin{figure}[ht] %ORIGINAL SIZE: width=1.4TRUEIN; height=1.5TRUEIN
\vspace*{5pt}
\centerline{\psfig{file=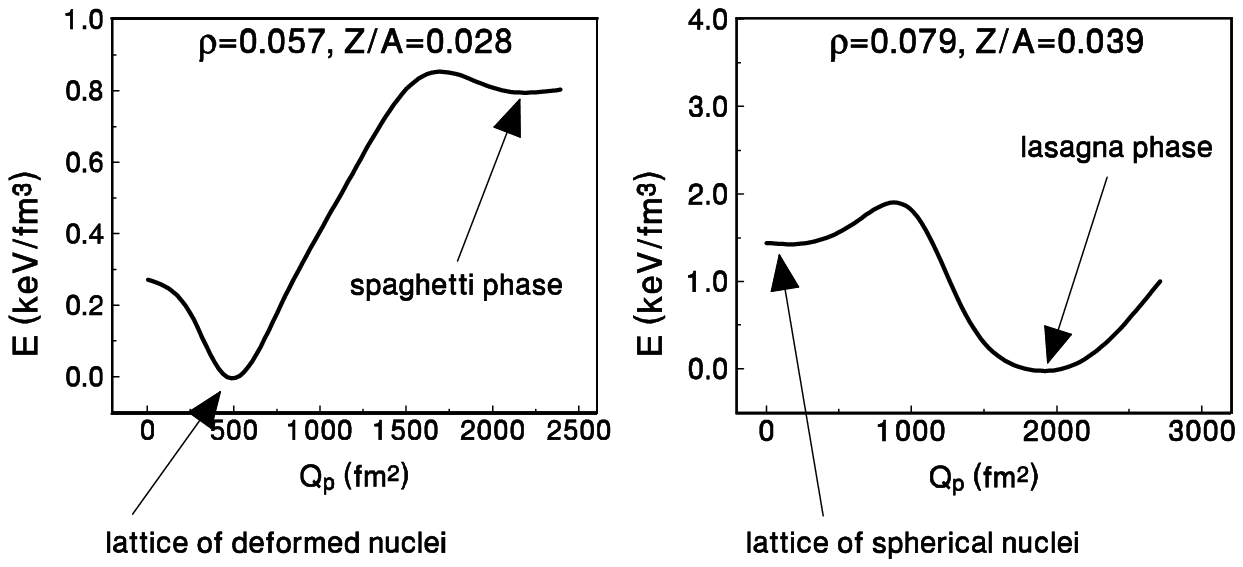}} %100 percent
\vspace*{5pt}
\fcaption{The total energy density (\ref{energy}) of
the npe matter as a function of the proton
quadrupole moment $Q_p=Q^{p}_{20}$, calculated
in the HF approximation at the constant nuclear density $\rho$,
and proton fraction $Z/A$, where $Z$ and $A$ are the proton
and nucleon numbers per cell, respectively. The lattice constant is equal
to $26$ fm (left subfigure) and $20.8$ fm (right subfigure).
The stable nuclear configurations are indicated by arrows.}
\end{figure}

The determination of the shell
energy requires the calculation of the scattering matrix
of the system \cite{bwi}. It is
in principle a difficult task for arbitrary shapes and
arbitrary mutual geometrical arrangements of inhomogeneities.
It is possible however to obtain an
approximate analytic result under simplifying assumptions concerning
the nuclear shape. Namely,
let us consider here only spherical, cylindrical and
planar nuclei. Moreover since the amplitude of the wave
function in the semiclassical limit is proportional to the inverse
square root of the local momentum, the single particle wave functions
for the unbound states will have a small amplitude over the deep
well and thus the probability to find an unbound neutron
inside the nuclei is reduced as compared
to the surrounding space\footnote{
There are of course a number of ``resonant''
delocalized states, whose amplitudes behave in an opposite manner.
However, the number of such
``resonant'' states is small and brings only small corrections.}.
Hence one can approximately replace the nuclear potential
by an effective repulsive potential of roughly the same
shape \cite{bcf}.
Then the shell energy can be easily determined in the semiclassical
approximation based on the Gutzwiller trace formula.
Namely, the leading shell energy contribution to
the total energy, for two obstacles being either spherical,
rod-like or slab-like nuclei located at a distance $d$
reads \cite{mph,mph1}:
\begin{equation} \label{shellapr}
E_{shell}^{out} \approx
\frac{\hbar^{2} L^{i} R^{2-i}}{8m_{n}} \left ( \frac{3}{\pi}
\right )^{\frac{2+i}{6}}
\frac{(\rho_{n}^{out})^{\frac{2+i}{6}}}{d^{\frac{6-i}{2}}}
\cos\left (2k^{n}_{F}d-i\frac{\pi}{4}\right ) ,
\end{equation}
where $m_{n}$ is the neutron
mass and $i=0,1,2$ for two spherical, cylindrical and planar obstacles,
respectively (we assume that rods and slabs are parallel to each other).
In the above equation $L$ defines the length of the obstacle, and $R$ is
its radius (in the case of a slab it is defined
as half of its width).

Clearly, the interaction induced
by shell effects depends on the nuclear shape and
is a sensitive function of both
the neutron density and the geometry of the mutual
arrangement of nuclei. The interaction between many nuclei
of various shapes
may look, at first glance, quite complicated
since three--, four--, and other
many--body terms will appear as a result
of multiple neutron scattering on inhomogeneities.
One can show however that many-body terms are quite small and
give merely a small corrections to the dominant pairwise
interaction \cite{bwi}.

The magnitude of the quantum corrections associated
with the expression (\ref{shellapr}) is small as compared to
the liquid-drop terms in eq. (\ref{energy}). What is important,
however, is that the energy differences between various nuclear
phases in the crust are also very small (of the order of 
a few $keV/fm^{3}$),
since the liquid drop terms almost cancel. One can see it in
the figure 1, where the energy density of npe matter has
been plotted as a function of nuclear deformation. The minima
visible in the figure represent the stable nuclear phases
corresponding to different shapes or lattice geometries.
The results which are presented were obtained through the
minimization of the total energy functional (\ref{energy})
for the npe matter. Namely,
we solved the Hartree-Fock equations for a fixed density
of the npe
matter with contributions in the energy functional coming from the nuclei,
the neutron gas, and the electrons. In such a calculation, the
liquid drop  and the shell energy parts of the energy are automatically
and self-consistently included.
Since we solve the problem in a cubic box with three symmetry planes,
several lattice geometries can be generated like e.g. simple
cubic crystal (scc), face centered crystal (fcc), or body centered
crystal (bcc) (see Refs. \cite{mph,bfh} for details).

Since the liquid drop energy for different phases is almost
the same thus the shell energy plays a crucial role for
determining the most energetically favored structure.
In particular, it can be shown that
the relative energies of different phases fluctuate rapidly
as a function of the total density
and these oscillations can be attributed to the shell effects
associated with unbound neutrons \cite{mph,mph1,bma1}.
Consequently, we may conclude that the
structure of the crust may be quite complicated since the shell
effects associated with  unbound neutrons
may easily reverse the phase transition order predicted by the
liquid drop based approaches.  Moreover the number of phase
transitions may increase since the same phase may appear
for various density ranges.
It is also
likely that the system will favor distorted lattices, or lattices
with defects which decrease the shell energy.

Hence our results
suggest that the purely quantum effect related to
the fluctuation induced interaction between nuclei, and which thus
may be termed the fermionic Casimir effect, is responsible
for determining the structure of the inner crust of neutron stars.

\nonumsection{Acknowledgements}
\noindent
This research was supported in part by the Polish Committee
for Scientific Research (KBN) under Contract No.~5~P03B~014~21
and the Wallonie/Brussels-Poland
integrated action program. Numerical calculations were performed
at the Interdisciplinary Centre for Mathematical and Computational
Modelling (ICM) at Warsaw University. Authors would like to thank
Hubert Flocard for providing us with the numerical code
solving the Poisson equation. We also appreciate many discussions
with A. Wirzba, Y. Yu, S.A. Chin and H. Forbert.

\nonumsection{References}

\end{document}